\documentclass[aps,twocolumn,showpacs,superscriptaddress,pra,10pt]{revtex4-1}
\usepackage{graphicx,braket,amsmath,amsfonts,hyperref,color}

\newcommand{\crt}[1]{\hat{a}^\dagger_{#1}}
\newcommand{\dst}[1]{\hat{a}^{\phantom{\dagger}}_{#1}}
\newcommand{\vett}[1]{{\bf{#1}}}

\newcommand*\chem[1]{\ensuremath{\mathrm{#1}}}

\begin{document}

\author{Mario Motta}
\affiliation{Department of Physics, College of William and Mary, Williamsburg, Virginia 23187-8795, USA}

\author{Shiwei Zhang}
\affiliation{Department of Physics, College of William and Mary, Williamsburg, Virginia 23187-8795, USA}

\title{
Computation of ground-state properties in molecular systems: \\
back-propagation with auxiliary-field quantum Monte Carlo
}

\begin{abstract}
We  address the computation of ground-state properties of chemical systems and
realistic materials within the auxiliary-field quantum Monte Carlo method. The
phase constraint to control the fermion phase problem requires the random walks
in Slater determinant space to be open-ended with branching.
This in turn makes it necessary to use back-propagation (BP) to compute
averages and correlation functions of operators that do not commute with the Hamiltonian.
Several BP schemes are investigated and their optimization with respect to the
phaseless constraint is considered.
We propose a modified BP method for the computation of observables in electronic
systems, discuss its numerical stability and computational complexity, and assess
its performance by computing ground-state properties for several substances,
including constituents of the primordial terrestrial atmosphere and small organic
molecules.
\end{abstract}

\maketitle

\section{Introduction}

The quantitative study of molecular systems requires accurate and efficient calculations 
of ground--state properties and correlation functions.
For many applications, this task corresponds to solving the electronic many-body 
Schr\"odinger equation within the Born-Oppenheimer approximation, and computing observables 
and correlation functions as ground-state averages of suitable operators.

The last decades have witnessed the development of a variety of numerical techniques to 
address the solution of the many-body Schr\"odinger equation, ranging in quality from the 
mean-field Hartree-Fock (HF) to the exact full configuration interaction (FCI).
The coupled-cluster (CC) and quantum Monte Carlo (QMC) methods 
\cite{Szabo1996,Bartlett2007,Hammond1994,Foulkes2001} are two examples of techniques that 
incorporate electronic correlations beyond the HF level to achieve a better description 
of the system.

QMC methods provide a non-perturbative treatment of interacting many-body systems, typically 
with excellent accuracy.
Their computational complexity scales as a low power of the system size. With the advent of 
powerful computational resources, they are becoming increasingly valuable for unveiling the 
properties of quantum many-particle models and real materials 
\cite{Foulkes2001,Esler2008,Motta2017}.

A broad class of projector QMC methods are based on the mapping between the imaginary-time 
evolution and a suitable stochastic process \cite{Hammond1994}. Projector QMC methods differ 
from each other by the choice of the mapping and by the approximation schemes employed to 
control the well-known sign problem \cite{Loh1990}.
In particular the QMC method with the longest history in electronic structure is the real 
space fixed-node diffusion Monte Carlo (DMC), which samples the many-body ground-state 
function in the configurational space of the studied system 
\cite{Reynolds1982,Hammond1994,Foulkes2001}.
 
The more recent auxiliary-field quantum Monte Carlo (AFQMC) method is an alternative and 
complementary QMC approach, based on the mapping between the imaginary-time evolution and 
a stochastic process taking place in a non-orthogonal manifold of Slater determinants 
\cite{Zhang2003}. 
AFQMC has provided excellent results for ground-state energies and energy differences in 
solids and molecules 
\cite{AlSaidi2006,AlSaidi2007,Suewattana2007,Purwanto2008,Purwanto2009,Purwanto2011,Purwanto2016,Shee2017}.
While AFQMC has been extensively applied in lattice models to calculate both ground-state 
energies and other observables and correlation functions \cite{LeBlanc2015,Qin2016,Zheng2017}, 
the calculation of ground-state properties other than the total energy has to date not 
been systematically investigated in molecules and other realistic materials.
 
In the present work, we formulate and implement a back-propagation (BP) technique 
\cite{Zhang1997,Purwanto2004} for calculating ground-state properties of realistic electronic 
Hamiltonians with the AFQMC method.
We show that, under the formalism adopted in electronic systems, which often involves a 
phaseless constraint, a simple refinement of the original back-propagation \cite{Purwanto2004} 
leads to substantial improvement in the quality of the computed observables.
The theoretical basis for the improvement is discussed.
We then give a detailed description of our algorithm, and consider its numeric stability 
and computational complexity. To assess the accuracy of the methodology, we compute ground-state 
averages of one-body operators and two-body correlation functions for several molecules, 
including constituents of the primordial terrestrial atmosphere and small organic molecules.
Results are compared with FCI and CC where available.

The remainder of this paper is organized as follows. 
The AFQMC method for total energy calculations (without back-propagation) is outlined in 
Sec.~\ref{sec:afqmc-method}. The back-propagation technique and our implementation as well as 
technical improvements for electronic systems are then described in Sec.~\ref{sec:methods}. 
Results are presented in Sec.~\ref{sec:results} and conclusions are drawn in Sec.~\ref{sec:conclusions}.

\section{The AFQMC method}

\label{sec:afqmc-method}

We first provide a brief overview of the AFQMC method \cite{Zhang2003,Zhang2013}, to establish 
the notation and present the concepts relevant to the formulation of the back-propagation algorithm.
The hamiltonian of a many-electron system with spin-independent external potential and two-body 
interactions can be written as
\begin{equation}
\label{eq:the_hamiltonian}
\begin{split}
\hat{H} &= \sum_{ij,\sigma} T_{ij} \, \crt{i\sigma} \dst{j\sigma} + \sum_{ijkl,\sigma\tau}^M 
\frac{V_{ijkl}}{2} \, \crt{i\sigma} \crt{j\tau} \dst{k\tau} \dst{l\sigma} = \\
&= \hat{H}_1 + \hat{H}_2
\end{split}
\end{equation}
where the creation and annihilation operators $\crt{i\sigma}, \, \dst{j\sigma}$ are related to 
an orthonormal set of one-electron orbitals $ \{ \varphi_i({\bf{r}}) \}_{i=1}^M$, and the 
one-body and two-body matrix elements $T_{ij}$, $V_{ijkl}$ are generated by quantum chemistry 
softwares (or via a downfolding procedure using Kohn-Sham orbitals \cite{Ma2015}).

The AFQMC method is based on the observation that the ground state $\Psi_0$ of 
\eqref{eq:the_hamiltonian} can be expressed as 
\begin{equation}
\ket{\Psi_0} = \lim_{\beta\to\infty} e^{-\beta (\hat{H}-E_0)} \ket{\Psi_I}
\end{equation}
where $\Psi_I$ is any initial wavefunction having non-zero overlap with $\Psi_0$, and $E_0$ the 
ground-state energy of $\hat{H}$. (In an AFQMC calculation $E_0$ is replaced by a trial energy 
which can be estimated adaptively.)
Applying the propagator $e^{-\beta (\hat{H}-E_0)}$ to $\Psi_I$ requires the discretization of 
the imaginary-time propagation
\begin{equation}
\label{eq:hs1}
e^{-\beta (\hat{H}-E_0)} = \left( e^{-\delta\beta (\hat{H}-E_0)}  \right)^n 
\,, \quad 
\delta\beta = \frac{\beta}{n}\,,
\end{equation}
and the combined use of the Trotter-Suzuki decomposition \cite{Trotter1959,Suzuki1976}
\begin{equation}
\label{eq:hs2}
e^{-\delta\beta \hat{H}} = e^{-\frac{\delta\beta}{2} \hat{H}_1} e^{-\delta\beta \hat{H}_2} 
e^{-\frac{\delta\beta}{2} \hat{H}_1} + \mathcal{O}(\delta\beta^3)
\end{equation}
and the Hubbard-Stratonovich transformation \cite{Stratonovich1957,Hubbard1959}
\begin{equation}
\label{eq:hs3}
e^{-\delta\beta (\hat{H}-E_0)} = \int d\vett{x} \, p(\vett{x}) \, \hat{B}(\vett{x}) + 
\mathcal{O}(\delta\beta^2)\,.
\end{equation}
In \eqref{eq:hs3}, the operators $\hat{B}(\vett{x})$ are given by
\begin{equation}
\label{eq:Bx}
\hat{B}(\vett{x}) = e^{-\frac{\delta\beta}{2} \hat{H}_1} \prod_\gamma e^{\sqrt{\delta\beta} 
x_\gamma \hat{v}_\gamma} e^{-\frac{\delta\beta}{2} \hat{H}_1} 
\, e^{\delta\beta E_0}\,,
\end{equation}
the operators $\hat{v}_\gamma$, typically obtained through a modified Cholesky decomposition 
of the two-body tensor $V_{ijkl}$ \cite{Purwanto2011}, 
are such that $\hat{H}_2 = - \frac{1}{2} \sum_\gamma \hat{v}_\gamma^2$, and $p(\vett{x})$ is 
the normal probability distribution.
Equations \eqref{eq:hs1}, \eqref{eq:hs2} and \eqref{eq:hs3} yield the following stochastic 
representation of the ground-state wavefunction:
\begin{equation}
\label{eq:fp}
e^{-n \delta\beta (\hat{H}-E_0)} \ket{\Psi_I} \simeq \int d\vett{X} \, p(\vett{X}) \ket{ \Psi_n(\vett{X}) }
\end{equation}
where $\vett{X} = (\vett{x}_{n-1} \dots \vett{x}_0)$ is a path of auxiliary fields and 
$\ket{ \Psi_n(\vett{X}) } = \hat{B}(\vett{x}_{n-1}) 
\dots \hat{B}(\vett{x}_{0}) \ket{\Psi_I}$. We will frequently write $\Psi_n$ in place of 
$\Psi_n(\vett{X})$ to reduce clutter.

The practical usefulness of \eqref{eq:fp} rests upon Thouless's theorem, ensuring that the 
operators $\hat{B}(\vett{x}_{i})$, acting on a Slater determinant $\Psi_I$, yield another 
Slater determinant $\Psi_n$ \cite{Zhang2013}.
Equation \eqref{eq:fp} maps the imaginary-time projection onto a stochastic process in the 
manifold of Slater determinants, which can be efficiently simulated thanks to the well-known 
algebraic properties of these many-body states \cite{Zhang2013}. Such mapping gives rise 
to an algorithmic procedure via the so-called free-projection AFQMC, with computational 
cost scaling as $\mathcal{O}(M^2 N + N^3)$ for propagation and possibly $\mathcal{O}(M^4)$ 
for computing two-body expectation values (for the energy, the scaling can be reduced to 
$\mathcal{O}(M^2 N^2)$ or better.)

\subsection{Importance sampling}
\label{ssec: imp-sampl}

The straightforward free-projection AFQMC manifests a well-known exponential increase in 
statistical fluctuations with the projection time $n \, \delta\beta$ \cite{Shi2013}. 
This is the sign problem \cite{Loh1990}, which for general Coulomb interactions turns into 
a phase problem \cite{Zhang2003}.
The problem is controlled by combining an importance sampling procedure, briefly overviewed 
below, with suitable approximations.

The starting point of the importance sampling is performing a shift of the auxiliary fields 
in \eqref{eq:fp}
\begin{equation}
\label{eq:is}
\resizebox{\columnwidth}{!}{$
\begin{split}
& e^{-n \delta\beta (\hat{H}-E_0)} \frac{ \ket{\Psi_I} }{ \braket{\Psi_T | \Psi_I} } \simeq 
  \int d\vett{X} \, p(\vett{X}-\overline{\vett{X}}) \, 
       \frac{ \ket{ \Psi_n(\vett{X}-\overline{\vett{X}}) } }{ \braket{\Psi_T | \Psi_I} } \\
& = \int d\vett{X} \, p(\vett{X}) \, W_n(\vett{X},\overline{\vett{X}}) \, 
       \frac{ \ket{ \Psi_n } }{ \braket{ \Psi_T | \Psi_n }  }
\end{split}
$}
\end{equation}
where $\Psi_T$ is a trial wavefunction used to guide the simulation, which can be different from $\Psi_I$ if desired, the vector $\overline{\vett{X}}$ can be complex-valued, and the weights $W_n(\vett{X},\overline{\vett{X}})$ are defined in terms of the importance function
\begin{equation}
I(\vett{x},\overline{\vett{x}}, \Psi ) = \frac{p(\vett{x}-\overline{\vett{x}})}{p(\vett{x})} \, \frac{ \braket{\Psi_T | \hat{B}(\vett{x}-\overline{\vett{x}})|\Psi} }{ \braket{\Psi_T|\Psi} }
\end{equation}
as
\begin{equation}
W_n(\vett{X},\overline{\vett{X}}) = \prod_{i=0}^{n-1} I(\vett{x}_i,\overline{\vett{x}}_i, \Psi_i ) 
\,.
\end{equation}
The components $\overline{\vett{x}}_i$ of $\overline{\vett{X}}$ are chosen \cite{Purwanto2004} to minimize fluctuations in the importance function to first 
order in $\delta\beta$, and read
\begin{equation}
\label{eq:xbar-choice}
\left( \overline{\vett{x}}_i \right)_{\gamma} = - \sqrt{\delta\beta} \frac{ \braket{ \Psi_T | \hat{v}_\gamma | \Psi_i }  }{ \braket{ \Psi_T | \Psi_i }  } \,.
\end{equation}
Although the vectors $\overline{\vett{x}}_i$ drive the random walk towards the region of the complex plane where $\braket{ \Psi_T | \Psi}$ is large,  the importance sampling, which is a similarity transformation, is not able to stabilize the random walk of Slater determinants.
The phase problem manifests itself in two aspects after importance sampling.
First, walkers whose overlap with $\Psi_T$ is small in magnitude cause large fluctuations in the weights and their contributions to the estimators.
Second, the weights in \eqref{eq:is} are complex and these quantities diffuse in the complex plane, resulting in cancelling signals in the Monte Carlo 
estimators of ground-state properties and correlation functions \cite{Zhang2013}.

\subsection{The phaseless approximation}

To achieve complete control of the phase problem, we rely on a phaseless approximation, where the importance function is modified as follows:
\begin{equation}
\label{eq:rle}
I(\vett{x},\overline{\vett{x}}, \Psi ) \simeq e^{ \delta\beta \left[ E_0 - \mathrm{Re}(E_L)\right] }
\times 
\max\left( 0, \cos(\Delta\theta) \right) \,,
\end{equation}
the local energy $E_L$ and $\Delta\theta$ being
\begin{equation}
\begin{split}
\label{eq:El-def}
E_L(\Psi) = \frac{ \braket{ \Psi_T | \hat{H} | \Psi} }{ \braket{ \Psi_T | \Psi} }\quad ,  \\
\Delta\theta = \mbox{Arg} 
\frac{ \braket{ \Psi_T | \hat{B}(\vett{x}-\overline{\vett{x}}) | \Psi}  }
     { \braket{ \Psi_T | \Psi} } \quad .
\end{split}
\end{equation}
The first factor
in Eq.~\eqref{eq:rle}, corresponding to the real local energy approximation, turns weights into real and positive quantities.
The second, corresponding to a projection after the proper gauge condition 
has been imposed with the choice of 
$\overline{\vett{x}}$ in Eq.~(\ref{eq:xbar-choice}), prevents the scalar products $\braket{ \Psi_T | \Psi}$ from undergoing 
a rotationally-invariant random walk in the complex plane, thus avoiding a finite concentration of walkers at the origin. 

In Eq.~(\ref{eq:rle}) the `$\cos$' can be replaced by a line constraint or a Gaussian weight \cite{Zhang2003,Zhang2005}.
These share the same basic idea as above and were seen to give similar results for the computed 
ground-state energy in the tests in jellium \cite{Zhang2003} and repulsive Bose systems with modest interactions \cite{Purwanto2005}.
Regarding the first part, 
it was found that the imaginary part of the local energy accumulates slowly and
can be carried for extended projection time in electronic systems; however this seems to have little 
effect on the computed ground state energy.

\subsection{The phaseless AFQMC algorithm for total energy}

The resulting algorithm, the phaseless AFQMC, can be summarized by the following sequence of operations:
\begin{enumerate}
\item $N_w$ walkers, labeled by $k$, are initialized at $\Psi_{0,k} \equiv \Psi_I$, each with weight $W_{0,k} \equiv 1$. (If a multi-determinant  $\Psi_I$ is used, the different determinants can be sampled 
according to their weights in  $\Psi_I$.)

\item For each $k$, $\vett{x}_{i,k}$ is sampled, the walker is updated as 
\begin{displaymath}
\Psi_{i+1,k} = \hat{B}((\vett{x}-\overline{\vett{x}})_{i,k}) \Psi_{i,k} \,
\end{displaymath}
and the weight as 
\begin{displaymath}
W_{i+1,k} = I(\vett{x}_{i,k},\overline{\vett{x}}_{i,k},\Psi_{i,k}) \, W_{i,k}
\end{displaymath}

\item Step 2 is iterated $n$ times, and Eq.~(\ref{eq:is}) is realized in a Monte Carlo sense as a weighted average as ${\mathcal N}\,\sum_{k=1}^{N_w}
W_{n,k}  \frac{ \ket{\Psi_{n,k} } }{ \braket{ \Psi_T | \Psi_{n,k} } }$, 
where ${\mathcal N}$ is a normalization constant that depends on the trial energy $E_0$ and 
the sum of the weights. 

\item Periodically, the walkers $\{ \Psi_{i,k}\}$ are stabilized, for example, using a modified Gram-Schmidt 
procedure \cite{Zhang2013}. Additionally, the weights can be reorganized using a branching algorithm \cite{Zhang2003}.

\end{enumerate}
Additional algorithmic improvements enhance the efficiency and stability of AFQMC calculations, for instance subtracting a mean-field contribution 
to the two-body part of the Hamiltonian prior to the HS transformation \cite{AlSaidi2006,Purwanto2009,Shi2013} and bounding weights and force biases \cite{Purwanto2009}. 
This approach was successfully applied to a broad number of quantum chemistry \cite{Purwanto2014,Virgus2014,Motta2017} and solid state physics systems \cite{Purwanto2004,Ma2015}. 

\section{Computation of observables and correlation functions in AFQMC}
\label{sec:methods}

The importance sampling transformation provides a stochastic representation of the ground-state wavefunction as a weighted 
average of rescaled Slater determinants, and gives the possibility of computing the mixed estimator
\begin{equation}
\mathrm{A_{mix}} = \frac{ \braket{ \Psi_T | \hat{A} | \Psi_0} }{ \braket{ \Psi_T | \Psi_0} }
\end{equation}
of an observable $\hat{A}$ as a weighted average
\begin{equation}
\mathrm{A_{mix}} \simeq 
\frac{1}{\sum_{k} W_{n,k} } \, 
\sum_{k} W_{n,k} \frac{ \braket{ \Psi_T | \hat{A} | \Psi_{n,k} } }{ \braket{ \Psi_T | \Psi_{n,k} } }  \,,
\end{equation}
where the rescaled matrix elements can be evaluated in a manner similar to the total energy.
Unless $[\hat{A},\hat{H}]=0$, the mixed estimator of $\hat{A}$ is biased by the trial wavefunction $\Psi_T$ used for 
importance sampling. A simple method to approximately remove the bias in the mixed estimator is
the extrapolation
\begin{equation}
\mathrm{A_{ex}} = 2 \mathrm{A_{mix}} -  \mathrm{A_{T}} 
\,\, , \,\,
\mathrm{A_{T}} = 
\frac{
\braket{ \Psi_T | \hat{A} | \Psi_T} 
}{
\braket{ \Psi_T | \Psi_T} 
} 
\,\, .
\end{equation}
Expectation values obtained through the extrapolated estimator, however, 
still contain bias from the choice of the trial wavefunction
which is difficult to assess. 
Indeed, because the trial wave functions tend to be quite poor (often a single Slater determinant) in AFQMC, the extrapolated
estimator can often be worse than the mixed estimator \cite{Purwanto2004}.

\subsection{Phaseless back-propagation (BP-PhL)}

\label{subsec:simple-bp}

In order to overcome these restrictions, Zhang and coworkers proposed a back-propagation (BP) technique \cite{Zhang1997,Purwanto2004} 
in the framework of AFQMC. The starting point of the BP algorithm is the observation that, for large $n$ and $m$,
\begin{equation}
\label{eq:bp1}
\mathrm{A_{BP}} \equiv
\frac
{ 
\braket{ \Psi_T | e^{-m \delta\beta \hat{H}} \hat{A} e^{-n \delta\beta \hat{H}} | \Psi_I} 
}{ 
\braket{ \Psi_T | e^{- (m+n) \delta\beta \hat{H}} | \Psi_I} 
}
\simeq
\frac{ \langle \Psi_0 | \hat{A} | \Psi_0 \rangle }{ \langle \Psi_0 | \Psi_0 \rangle }\,.
\end{equation}
The measurement above reduces to the mixed estimator when $m=0$.
Inserting \eqref{eq:fp} into \eqref{eq:bp1} 
yields
\begin{equation}
\label{eq:expectationBP}
\mathrm{A_{BP}} = 
\frac
{ 
\int d\vett{X} \, p(\vett{X}) \braket{ \Phi_{m}(\vett{X}) | \hat{A} | \Psi_n(\vett{X}) }
}{ 
\int d\vett{X} \, p(\vett{X}) \braket{ \Phi_{m}(\vett{X}) |      \Psi_n(\vett{X}) }
}
\end{equation}
with $\vett{X} = (\vett{x}_{n+m-1} \dots \vett{x}_0)$ and 
\begin{equation}
\label{eq:Bprod-bp}
\ket{ \Phi_{m}(\vett{X}) } = \hat{B}(\vett{x}_{n})^\dagger \dots \hat{B}(\vett{x}_{n+m-1})^\dagger \ket{ \Psi_T }\,.
\end{equation}
As before, we will use the abbreviation $\Phi_m$ to denote $\Phi_{m}(\vett{X})$.
Applying  the shift on the path segment from $n+1$ to $m$ in the forward direction, i.e., 
viewing the  entire path  $\vett{X}$ as a forward projection from $|\Psi_I\rangle$  and 
performing the importance sampling transformation of Sec.~\ref{ssec: imp-sampl}, yields
\begin{equation}
\label{eq:bp_formula}
\mathrm{A_{BP}} = 
\frac
{ 
\int d\vett{X} \, p(\vett{X}) \, W_{n+m}(\vett{X},\overline{\vett{X}}) 
\, 
\frac{ \braket{ \Phi_m | \hat{A} | \Psi_n } }
       { \braket{ \Phi_m |      \Psi_n } } 
}{ 
\int d\vett{X} \, p(\vett{X}) \, W_{n+m}(\vett{X},\overline{\vett{X}}) 
}\,.
\end{equation}
In a calculation with importance sampling, the quantity $\mathrm{A}_{\rm BP}$ is thus estimated as the following weighted average
\begin{equation}
\label{eq:bp}
\mathrm{A_{BP}} \simeq
\frac{1}{\sum_{k} W_{n+m,k} } \, 
\sum_{k} W_{n+m,k} \frac{ \braket{ \Phi_{m,k} | \hat{A} | \Psi_{n,k} } }{ \braket{ \Phi_{m,k} | \Psi_{n,k} } }\,,
\end{equation}
with
\begin{equation}
\label{eq:fwd-proj_dets}
\ket{ \Psi_{n,k} } = \hat{B}\left((\vett{x}-\overline{\vett{x}})_{n-1,k}\right) \dots \hat{B}\left((\vett{x}-\overline{\vett{x}})_{0,k}\right)  \ket{ \Psi_I }
\end{equation}
which is the usual forward projection in total energy calculations, and 
\begin{equation}
\label{eq:bp_dets}
\resizebox{\columnwidth}{!}{$
\ket{ \Phi_{m,k} } = \hat{B}^\dagger\left((\vett{x}-\overline{\vett{x}})_{n,k}\right) \dots \hat{B}^\dagger\left((\vett{x}-\overline{\vett{x}})_{n+m-1,k}\right) 
\ket{ \Psi_T }
$}
\end{equation}
which uses the auxiliary-field path in the backward direction for projection.
Note that the matrix elements in Eq.~\eqref{eq:bp} can be efficiently evaluated 
for both one- and two-body operators \cite{Zhang2013}.
One important advantage of the AFQMC method with BP is that the Monte Carlo samples,
$\{\langle \Phi_m|\}$  and  $\{ | \Psi_{n} \rangle\}$, are non-orthogonal, 
which allow essentially any correlation functions to be computed conveniently 
in electronic systems.

\begin{figure}[ht!]
\includegraphics[width=0.45\textwidth]{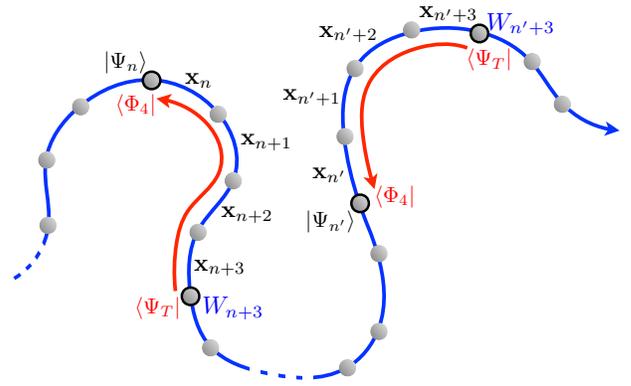}
\caption{(color online) Pictorial illustration of the back-propagation technique. 
For each walker $k$, $\ket{ \Psi_{n,k} }$ is computed at a time step $n$, $\ket{ \Phi_{m,k} } $ is computed between 
the time steps $n$ and $n+m$, and the weight $W_{n+m,k}$ at the time step $m+n$. 
Here the path for one walker is shown for two BP segments with
 $m=4$ back-propagation steps in each.
Evaluation of $\mathrm{A}_{BP}$ is performed periodically during the simulation following Eq.~\eqref{eq:bp}. 
} 
\label{fig:bp}
\end{figure}

The back-propagation technique, illustrated in Figure \ref{fig:bp}, makes use of auxiliary 
fields configurations from different segments of the random walk to project the 
trial wavefunction also at the left of $\hat{A}$. 
In the absence of any constraint for controlling the sign or phase problem,
this estimator approaches the exact expectation value as the number $m$ of back-propagation steps 
is increased.
When a constraint is applied in the forward direction, the backward paths do not satisfy 
the same constraint so that the BP estimators are approximate and not variational.

\subsection{Improved back-propagation algorithm - path restoration (BP-PRes)}
\label{sec:alg}

The BP approach described above  has been successfully applied to a variety of 
lattice models \cite{Shi2013,LeBlanc2015,Qin2016}, in which the local interactions induce ``only'' a sign problem
and the phaseless approximation reduces to the simpler constrained-path approximation of
$\langle \Psi_T|\Psi\rangle >0$. In the presence of a phase problem when the phaseless 
approximation is required, the simple approach of directly using the plaseless paths has been used in
weakly interacting boson gases \cite{Purwanto2005} and dense 
homogeneous electron gases \cite{Motta2014,Motta2015}. 

In model systems where the one-body propagator in Eq.~(\ref{eq:Bx}) is real (or only 
has a trivial phase uncoupled to the random fields \cite{Chang2008}), there is no ambiguity on
how to formulate the BP, since the only role of the constraint is to terminate a path in the forward direction
when the constraint is violated.
In the general case when $\hat B({\mathbf x})$ is complex,
there are additional modifications to the weights of the path (beyond the importance sampling transformation,
whose effect is properly accounted for by using the ``retarded'' weight in Eq.~\eqref{eq:bp}).
The importance function $I(\vett{x},\overline{\vett{x}}, \Psi )$ is equal to the exponential of the local energy,
 $e^{ \delta\beta \left( E_0 - E_L(\Psi) \right) }$,
up to terms of order $\delta\beta^{\frac{3}{2}}$ \cite{Motta2014}.
The real local energy approximation in Eq.~\eqref{eq:rle} removes the imaginary part of the local energy from the importance 
function, leaving positive weights. The cosine projection multiplies the importance function by another positive factor, to remove
 the ``two-dimensional''
nature of the random walk  (in the complex plane of overlaps).

The effect of the above two steps is to alter the weight in the forward direction. The accumulation of their 
effect can be to terminate certain paths; once this occurs it cannot be recovered in the backward direction. The 
numerical modification to the weight, however, can be restored for BP among the paths that survive 
from step $n$ to step $(n+m)$ in Eqs.~(\ref{eq:bp_formula}) and (\ref{eq:bp}). 
Given that the backward direction 
would be exact if no phaseless projection on the weight had been applied in the forward direction
(although this obviously would re-introduce the phase problem), it is reasonable to 
restore the paths as much as possible  in the backward direction, by 
undoing  the extra weight modifications that arose from the phaseless approximation.

We will call the simple BP procedure without undoing any weights BP-PhL (phaseless), 
and the procedure which restores the weight projection and the phase of the local energy in the BP direction
BP-PRes (path restoration).
This distinction only exists when there is a genuine phase problem from a 
complex $\hat B({\mathbf x})$ with phase factors coupled to ${\mathbf x}$. 
(As mentioned, when $\hat B({\mathbf x})$ is real, both of these approaches will revert back to the standard BP
formulated for the case of the sign problem \cite{Zhang1997}.)
In weakly interacting 
systems, the two approaches are expected to be similar, since the effect of the phase projection is weak. 

In the BP-PRes algorithm, we replace the BP estimator with
\begin{equation}
\label{eq:cosinephase}
\mathrm{A_{BP}} \simeq
\frac{1}{\sum_{k} \tilde{W}_{n+m,k} } 
\sum_{k} \tilde{W}_{n+m,k} \frac{ \braket{ \Phi_{m,k} | \hat{A} | \Psi_{n,k} } }{ \braket{ \Phi_{m,k} | \Psi_{n,k} } }\,,
\end{equation}
where
\begin{equation}
\tilde{W}_{n+m,k} = W_{n,k} \prod_{j=n}^{n+m} e^{ \delta\beta \left( E_0 - E_L (\Psi_j) \right) }\,.
\end{equation}
This is realized by 
remembering the phase factor from Im$(E_L)$ and the cosine projection 
factor in Eq.~\eqref{eq:rle} along the path from step $n$ to step $(n+m)$ for each walker, and 
multiplying $W_{n+m,k}$ by the former and dividing it by the latter, to give $\tilde{W}_{n+m,k}$
for the weight in the BP estimator in Eq.~\eqref{eq:cosinephase}.
As we illustrate in the next Section, this modification leads to significant improvement.

We stress that it is only in the backward direction that the restoration of the path occurs, i.e.,
retaining the imaginary part of the local energy and undoing the cosine projection. 
This procedure  increases the fluctuation in the BP estimators. However,
the BP algorithm is only applied 
for a finite 
time $\beta_{\rm BP}=m\,\delta\beta$, which must be kept as small as possible to avoid population instability, i.e., the 
population at step $(n+m)$ should come from ancestors which make up a significant of the population 
at step $n$.
As such, undoing the phase projection 
does not cause fundamental instabilities in the algorithm. As mentioned, 
the phase in the local energy can typically 
be carried for much longer than $\beta_{\rm BP}$ without noticeable effect. The restoring of 
the projection is only for the paths that survive in the forward direction and thus the additional 
weight factors are well regulated.

\subsection{Additional algorithmic discussions}

\subsubsection{Numeric stabilization and computational cost}

Periodic stabilization of the random walkers is necessary in the propagation in the forward direction
to prevent the single-particle orbitals in the Slater determinant from losing orthogonality 
from numerical noise (collapsing to a 
bosonic state). This is often done with a 
modified Gram-Schmidt decomposition, $\Psi=UDV$, where $U$ is an $M \times N$ matrix which 
contains orthonormal orbitals, $D$ is an $N\times N$ diagonal matrix, and $V$ is an upper triangular
matrix whose diagonal elements are $1$. 

Similarly, in the backward direction, when we propagate the trial wave function by 
Eq.~\eqref{eq:Bprod-bp} or  \eqref{eq:bp_dets}, stabilization is necessary. 
This can be handled straightforwardly following the same procedure as in the forward direction. 
Assuming that $\Psi_T$ is a Slater determinant (the case of multideterminants is discussed below),
the application of the propagators $\hat{B}^\dagger$, in reverse order starting from the $(n+m)$-th time slice
as indicated by Eq.~\eqref{eq:bp_dets}, leads to a new Slater determinant  $\Phi_{m,k}$.
We can stabilize  $\Phi_{m,k}$ periodically until the desired value of $m$ is reached, writing it in the $UDV$ form and discarding $D$ and $V$. 
This is because only $\Phi_{m,k}$ in orthonormal form is needed in 
the final estimator in Eq.~\eqref{eq:bp_dets},
while the contribution of $D$ and $V$ is already properly accounted for in the use of the
``retarded weight'' $W_{n+m,k}$ or $\tilde{W}_{n+m,k}$.
 
The computational cost of the calculation of $\mathrm{A}_{\rm BP}$ is thus of $\mathcal{O}(m M^2 N)$ operations, to compute 
the determinant at the bra of \eqref{eq:bp}, of $\mathcal{O}(M^2 N)+\mathcal{O}(M N^2)+\mathcal{O}(N^3)$ operations to compute the one-body 
Green's function in \eqref{eq:bp_dets}, and of $\mathcal{O}(M^2)$ operations to compute the rescaled matrix elements in \eqref{eq:bp}:
the additional cost of each BP segment of length $m$ in imaginary time 
is approximately the same as propagating the determinants in the forward direction for $m$ steps.
In most implementations (as is the case in the present work), the auxiliary-field configurations are stored 
and carried during the forward direction.
 Hybrid procedures are possible where one stores 
segments of propagators during the forward direction \cite{Vitali2016}, which would
reduce the amount of computing but possibly with tradeoff on the memory requirement.

\subsubsection{Free projection in the backward direction}

As mentioned, the BP time $\beta_{\rm BP}$ is typically modest, in order to ensure sufficient number of independent paths 
in the BP direction. Clearly it is possible to increase the overall population size to prolong $\beta_{\rm BP}$,
however the decay in the number of independent ancestry paths is rapid and the procedure is asymptotically 
unstable for $\beta_{\rm BP} \rightarrow \infty$. 
(This is in common with the  forward walking scheme \cite{Liu1974},
on which the basic idea of BP is based.)
We have not  found this to be a serious limitation, in 
a variety of AFQMC BP calculations in strongly correlated models. 

It is then reasonable to consider using a free-projection in the BP segment of the calculation. 
In other words, the imaginary-time propagation 
in the forward direction (i.e. at the ket) of Eq.~\eqref{eq:bp_formula}
is performed under the usual phaseless approximation to time $n$, and under free-projection 
from $n$ to $n+m$ so as to allow full path sampling in the backwards direction. 
Practically:
\begin{enumerate}
\item a large number of walkers is equilibrated under the phaseless constraint
\item the constraint is released going forward for some imaginary-time segment, and back-propagation is performed on this segment.
\item points 1,2 are iterated until a desired statistical accuracy is reached.
\end{enumerate}
The method brings back the phase problem, although there is a finite signal-to-noise ratio (in the spirit of 
a finite-temperature calculation). Its statistical errors will thus grow with  $N$ and $M$,
unlike in BP-PhL and BP-PRes calculations, which have well-behaved polynomial computational scaling.
The different BP algorithms explored in the present work are summarized in \ref{tab:1}

\begin{table}[t!]
\resizebox{\columnwidth}{!}{
\begin{tabular}{lcc}
\hline\hline
algorithm          & forward propagation    & backward propagation    \\
                   & $| \Psi_{n,k} \rangle$ & $\langle \Phi_{m,k} |$  \\
\hline
BP-PhL             & constrained            & phaseless               \\
\hline                                      
BP-PRes            & constrained            & carry Im$(E_{L})$;      \\
                   &                        & undo cosine projection  \\
BP-PRes (partial)  & constrained            & carry Im$(E_{L})$       \\
\hline                                         
BP-FP              & constrained            & free projection         \\
\hline\hline
\end{tabular}
}
\caption{Summary of the back-propagation algorithms employing
phaseless (PhL), path restoration (PRes), and 
free-projection (FP). In all cases, forward propagation is constrained.
Backward propagation is carried out by removing, at different levels,
the constraints applied in the forward direction. 
}
\label{tab:1}
\end{table}

\subsubsection{Multideterminant trial wavefunctions}
\label{subsubsec:multidet-psiT}

For total energy calculations, the phaseless approximation has demonstrated rather weak dependence
on the trial wave functions. Typically single-determinant $\ket{ \Psi_T }$ has been used. In strongly 
correlated and/or strongly multi-reference systems, the use of multideterminants as trial wavefunctions 
can improve the quality and efficiency of AFQMC calculations \cite{AlSaidi2007,Purwanto2009b,Purwanto2014,Purwanto2016}.

Formally it is straightforward to generalize the use of multideterminant trial wave function to our BP 
schemes. 
If the trial wavefunction is a linear combination
\begin{equation}
\ket{ \Psi_T } = \sum_\alpha A_\alpha \ket{ \Psi_T^{(\alpha)} }
\end{equation}
of Slater determinants $\Psi_T^{(\alpha)}$, 
the state $\Phi_{m,k}$ takes the form $\ket{ \Phi_{m,k} } = \sum_\alpha A_\alpha \ket{ \Phi^{(\alpha)}_{m,k} }$ with
\begin{equation}
\resizebox{\columnwidth}{!}{$
\ket{ \Phi^{(\alpha)}_{m,k} } = \hat{B}^\dagger\left((\vett{x}-\overline{\vett{x}})_{n,k}\right) \dots \hat{B}^\dagger\left((\vett{x}-\overline{\vett{x}}\right)_{n+m-1,k}) \ket{ \Psi_T^{(\alpha)} }
$}
\end{equation}
The stabilization procedure only requires minor modification to account for the linear combination.
Each determinant $\Phi^{(\alpha)}_{m,k}$ can be  stabilized separately.
The procedure yields a determinant $\overline{\Phi}^{(\alpha)}_{m,k}$, which comes from the $U^{(\alpha)}$
term in the modified Gram-Schmidt decomposition. 
However, unlike
in the case of a single determinant, the diagonal matrix $D^{(\alpha)}$ cannot be discarded, since it contributes to 
the relative weight:
\begin{equation}
\ket{ \Phi^{(\alpha)}_{m,k} } = \lambda^{(\alpha)}_{m,k} \ket{ \overline{\Phi}^{(\alpha)}_{m,k} }\,,
\end{equation}
i.e., the coefficient $ \lambda^{(\alpha)}_{m,k} $ should contain the  determinant of $D^{(\alpha)}$. Correspondingly, Eq.~\eqref{eq:bp_dets} takes the form
\begin{equation}
\frac{ \braket{ \Phi_{m,k} | \hat{A} | \Psi_{n,k} } }{ \braket{ \Phi_{m,k} | \Psi_{n,k} } }
= 
\frac{ 
\sum_\alpha \xi_{n,m,k}^{(\alpha)}
\, 
\frac{ \braket{ \overline{\Phi}^{(\alpha)}_{m,k} | \hat{A} | \Psi_{n,k} } }
     { \braket{ \overline{\Phi}^{(\alpha)}_{m,k}           | \Psi_{n,k} } }
}{
\sum_\alpha \xi_{n,m,k}^{(\alpha)}
}
\end{equation}
where $\xi_{n,m,k}^{(\alpha)} = A_\alpha \lambda^{(\alpha)}_{m,k} \braket{ \overline{\Phi}^{(\alpha)}_{m,k} | \Psi_{n,k} }$, and each of the mixed Green's functions can be computed as in Eq.~\eqref{eq:bp_dets}.
Furthermore, to prevent numeric overflows or underflows, the amplitudes $\xi_{n,m,k}^{(\alpha)}$
can be periodically rescaled by a common  factor.

The use of multideterminant trial wave functions offers a systematic route to improve the quality of the mixed and BP results. 
The major drawback of the multi-determinant trial wave function is the so-called size extensivity: to treat the electron correlation effects consistently, the number 
of Slater determinants required in a typical CI expansion grows rapidly with system size. For solid-state systems, 
it is typically impractical while,
in molecular systems,
this is less of a problem. When it is computationally feasible to generate a multi-determinant trial wave function
giving a good description of the molecule, for example with a truncated complete active space self-consistent field (CASSCF) calculation, 
the computational cost to use it in AFQMC only grows linearly with the number of Slater determinants in the trial wave function.
The corresponding efficiency can be even higher \cite{AlSaidi2007}, since a better trial wave function can reduce the 
statistical error as well as the systematic error.

\section{Results}
\label{sec:results}

We apply the BP algorithm to several atomic and molecular systems, primarily chosen for benchmarking purposes.
Most of the AFQMC calculations reported below used a single Slater determinant from HF as trial wave function.
It will be explicitly stated whenever additional tuning of the trial wave function beyond mean-field was performed.
We interfaced our AFQMC code with the NWChem \cite{Valiev2010} and PySCF \cite{Sun2016} quantum chemistry libraries to 
import the Gaussian one-electron and two-electron matrix elements, the overlap matrix, the trial wave function and the
atomic orbitals. 
All of our calculations are done using the spherical harmonics representation of basis functions.

For comparison, 
 we also employed well-established quantum chemistry methods where possible, including HF, 
second-order M{\o}ller-Plesset perturbation theory (MP2), CC, CASSCF, and FCI.
We performed CASSCF and FCI calculations using PySCF, and MP2 and CC calculations with NWChem.
Coupled-cluster calculations are of the type RCCSD, i.e., based on the restricted HF (RHF) reference state.

\subsection{Illustrative results and benchmark study}

\begin{figure}[h!]
\centering
\includegraphics[width=0.395\textwidth]{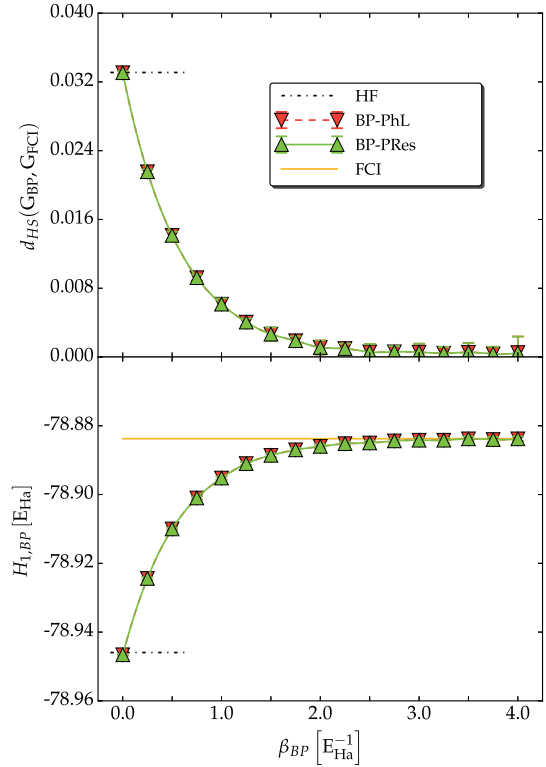}
\caption{
(color online) 
Evolution of the Hilbert-Schmidt distance between \chem{G_{BP}} and \chem{G_{FCI}}, and of the one-electron energy \chem{H_{1,BP}} (bottom) with 
back-propagation time for \chem{CH_4} (STO-3G level, tetrahedral geometry, $R_{\chem{CH}} = 1.1085$ \AA) using BP-PhL and BP-PRes.
} \label{fig:ch4}
\end{figure}

\begin{figure*}[t!]
\centering
\includegraphics[width=0.8\textwidth]{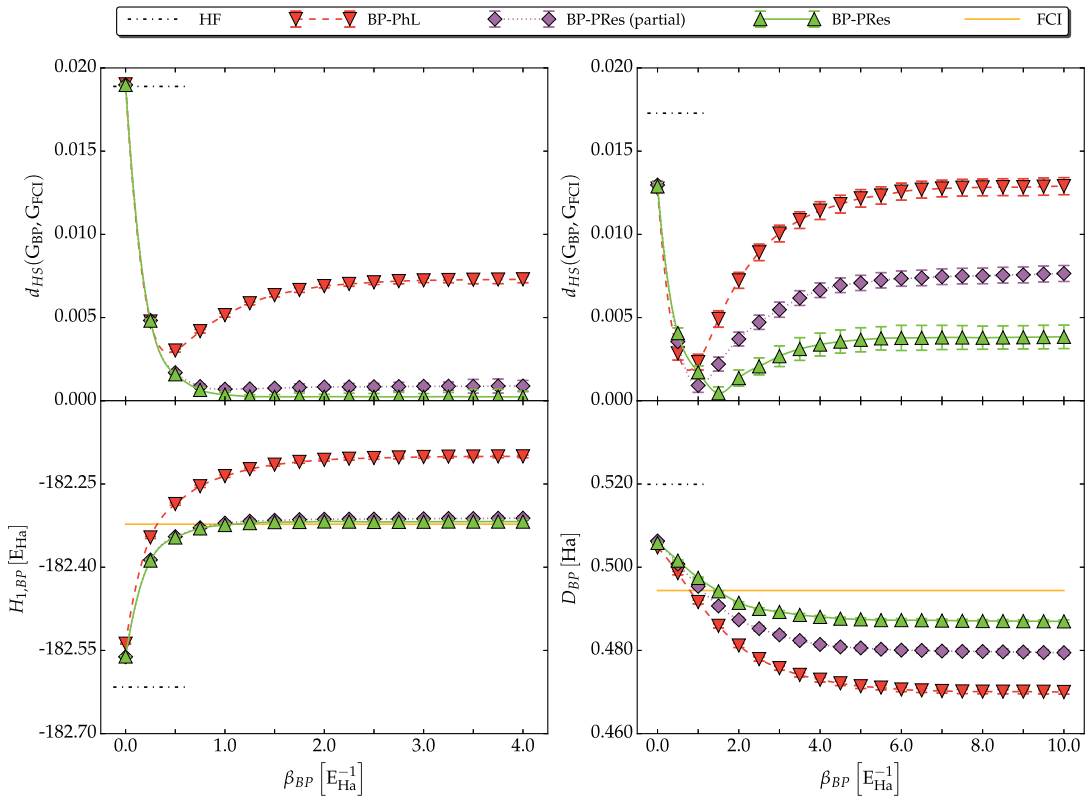}
\caption{
(color online)
Left: Evolution of the Hilbert-Schmidt distance between \chem{G_{BP}} and \chem{G_{FCI}} (top) and of the one-electron energy (bottom) with back-propagation 
time for \chem{Ne} (cc-pVDZ level) using PhL and PRes back-propagation.
Right: Evolution of the Hilbert-Schmidt distance between \chem{G_{BP}} and \chem{G_{FCI}} (top) and of the dipole moment (bottom) with back-propagation
time for \chem{HeH^+} (cc-pVDZ level, $R_{\chem{HeH}} = 0.79$ \AA) using BP-PhL and BP-PRes.
The results of a partial PRes (carrying the phase during BP but without undoing the cosine projection) 
are also shown.
} \label{fig:hehp}
\end{figure*}

\begin{figure}[h!]
\centering
\includegraphics[width=0.4\textwidth]{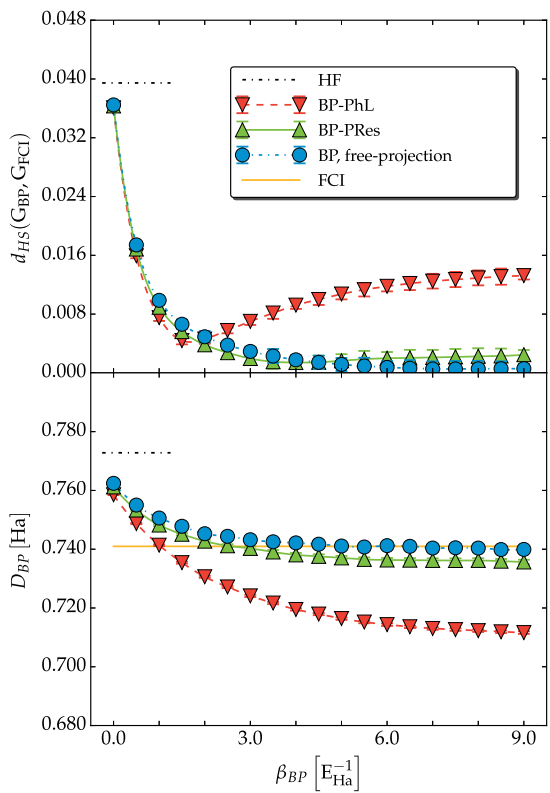}
\caption{
(color online)
Illustration of the BP-FP algorithm.
Evolution of the Hilbert-Schmidt distance between \chem{G_{BP}} and \chem{G_{FCI}} (top) and the 
dipole moment (bottom) is shown versus back-propagation time.
The system is \chem{NH_3} (STO-3G level, trigonal pyramid geometry, $R_{\chem{NH}} = 1.07$ \AA,
$\theta_{\chem{HNH}} = 100.08^o$).
} \label{fig:nh3}
\end{figure}

\begin{figure}[t!]
\centering
\includegraphics[width=0.4\textwidth]{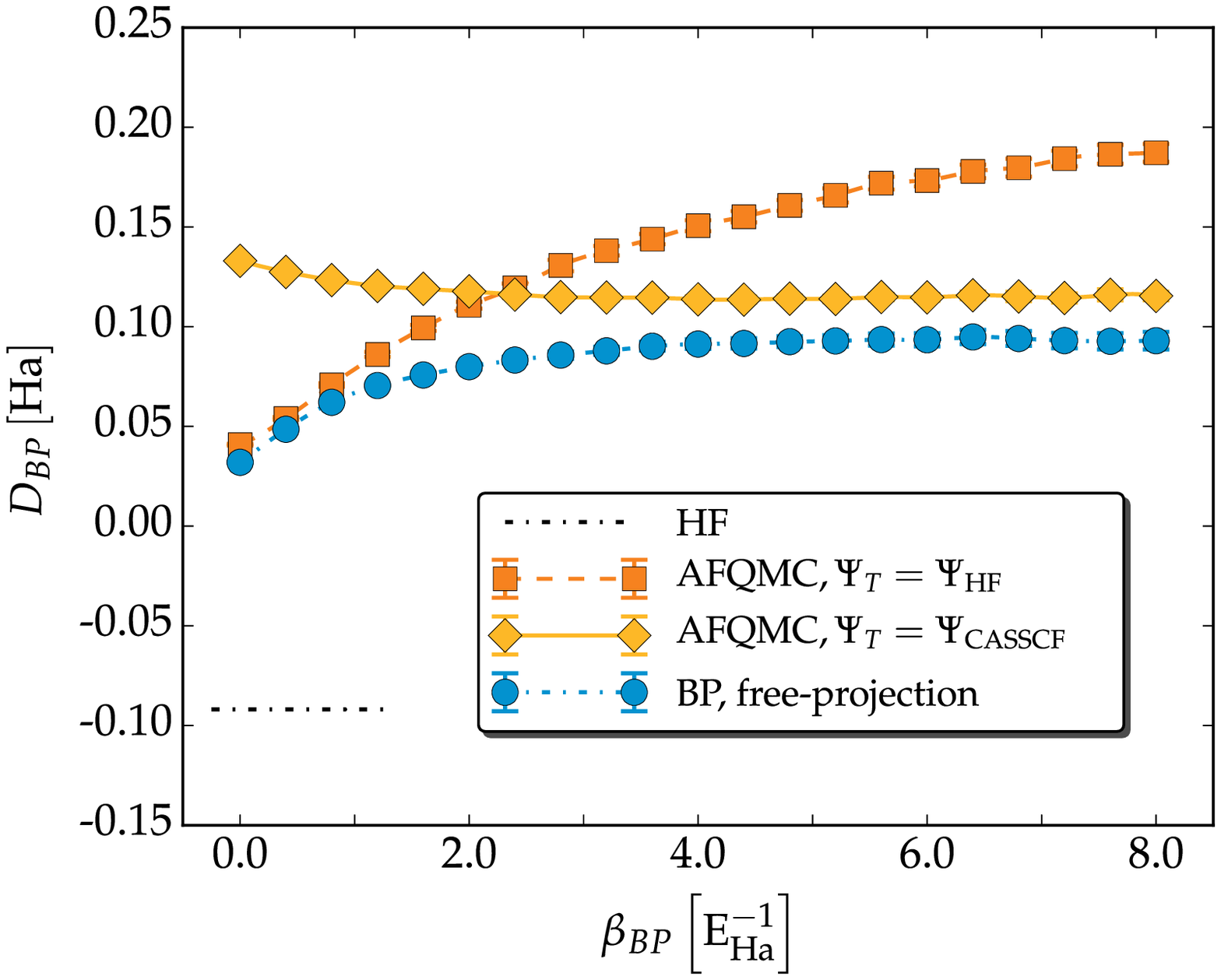}
\caption{
(color online)
Illustration of the use of multideterminant trial wave functions.
Evolution of the dipole moment with back-propagation time is shown 
for \chem{CO} (cc-pVDZ level, $R_{\chem{CO}} = 1.1282$ \AA) using 
 BP-PRes guided by  a truncated CASSCF wavefunction (yellow diamonds).
Results are compared to those from BP-FP (blue circles).
The corresponding  BP-PRes with RHF is also shown (orange squares) for reference.
} \label{fig:co}
\end{figure}

To study the accuracy of the BP algorithms, we compute the ground-state one-electron energy and dipole
moment for several small molecules using the STO-3G and cc-pVDZ basis sets \cite{Hehre1969,Dunning1989}.
The small number of electrons and atomic orbitals in these systems enable direct comparisons with FCI.
We also compute the spin-averaged one-body Green's function
\begin{equation}
\left( \mathrm{G_{BP}} \right)_{ij} = \frac{1}{2} \, \sum_{\sigma} 
\left\langle \crt{i\sigma} \dst{j\sigma} \right\rangle_{\mathrm{BP}} \,,
\end{equation}
and measure the error with respect to FCI by the Hilbert-Schmidt distance 
\begin{equation}
d_{HS}(\mathrm{G_{BP}},\mathrm{G_{FCI}})
=
\sqrt{ \mbox{Tr}\left[ (\mathrm{G_{BP}}-\mathrm{G_{FCI}})^2 \right] } \quad .
\end{equation}
In Figure \ref{fig:ch4} we show the computed Green's function and the one-electron energy in 
CH$_4$. The simple PhL back-propagation is seen to work well, leading to results in excellent agreement 
with FCI. Correspondingly, the ground-state energy computed from the mixed-estimator is -39.8069(1) $\mathrm{E_{Ha}}$,
compared to -39.8070 $\mathrm{E_{Ha}}$ from FCI. These observations are consistent with the system being 
weakly interacting and little bias being incurred by the
phaseless constraint imposed in the forward direction. 
As one would expect, 
the improved BP-PRes algorithm does not change the results from  PhL in this case.

Figure \ref{fig:hehp} illustrates the behaviors in a pair of less straightforward systems,
the Ne atom and the \chem{HeH^+} molecule. Here significant bias is seen 
from the  PhL algorithm. The computed ground-state energies from the mixed estimate 
are -128.6819(1) $\mathrm{E_{Ha}}$ (vs. -128.6809 from FCI) and -2.9612(1) $\mathrm{E_{Ha}}$
(vs -2.9609 $\mathrm{E_{Ha}}$ from FCI) respectively.
The BP-PRes algorithm leads to a systematic and substantial improvement in both cases.
To separate the contributions from the two parts in PRes, we also show here results from 
a partial restoration in the BP direction of the paths, by incorporating the complex phases
but omitting the restoration of the cosine projection. We see that, although the 
phase factor tends to play the more dominant role, both components can have a non-negligible 
effect.  This trend holds quite generally, and is consistent with the analysis in Sec.~\ref{sec:alg} 
on the rationale for
path restoration in the BP direction.

Tests conducted on 15 molecules at STO-3G, cc-pVDZ level show that the combination of the two adjustments in BP-PRes yields a systematic improvement 
over the PhL algorithm in all cases.
The average discrepancy in the computed one-electron energies and dipole moments is reduced by 
$\sim$ 50 \% on average, while the  Hilbert-Schmidt distances for the computed 
Green's functions is reduced by a factor of $\sim$ 3 on average.

We illustrate 
the effect of free-projection in the back-propagation direction (BP-FP) 
in Figure \ref{fig:nh3}. 
The BP-FP algorithm is seen to further reduce the small discrepancy from BP-PRes and yield essentially 
exact results on the Green's function and dipole moment. 
The  \chem{NH_3} example studied here is fairly challenging, as evidenced by 
the significant amount of bias shown by 
BP-PhL. The fact that BP-FP leads to very accurate results on the expectation values indicates
that the samples $ | \Psi_n(\vett{X}) \rangle$  in 
Eq.~(\ref{eq:expectationBP}) give an excellent representation
of the ground-state wave function, despite the phaseless constraint imposed 
when they are generated in the forward direction. This is consistent with the fact that the ground-state energies computed by phaseless AFQMC tend to be very accurate. 
The main source of inaccuracy in 
BP-PhL is in the backward direction, in which the paths constrained in the forward direction are 
a poor approximation when used in reversed direction.
The BP-PRes restores paths except for those terminated in the forward direction, hence the
significant improvement in accuracy.

The calculations discussed thus far have used the HF state as trial wavefunctions. 
It is straightforward to employ a multideterminant trial wavefunction as we have discussed in 
Sec.~\ref{subsubsec:multidet-psiT}.
An example is shown in Figure \ref{fig:co}.
The computation of the dipole moment of carbon monoxide is very difficult.
HF predicts the wrong  sign.
As a result,  AFQMC with BP-PRes using HF trial wave function still retains a significant error.
Using a truncated CASSCF(7,10) wavefunction as $\Psi_T$ brings the AFQMC estimate of the dipole 
moment much closer to the BP-FP result (the BP-FP calculation used a simple truncated 
CASSCF(3,7) trial wave function).
Convergence with $\beta_{\rm BP}$ is also improved, as well as the statistical accuracy. 
The improvement comes at the cost of a modest overhead, since the truncated CASSCF trial wavefunction 
in BP-PRes has a linear combination of 40 Slater determinants.

\subsection{Application to small and medium-sized molecules}

\begin{figure}[t!]
\centering
\includegraphics[width=0.94\columnwidth]{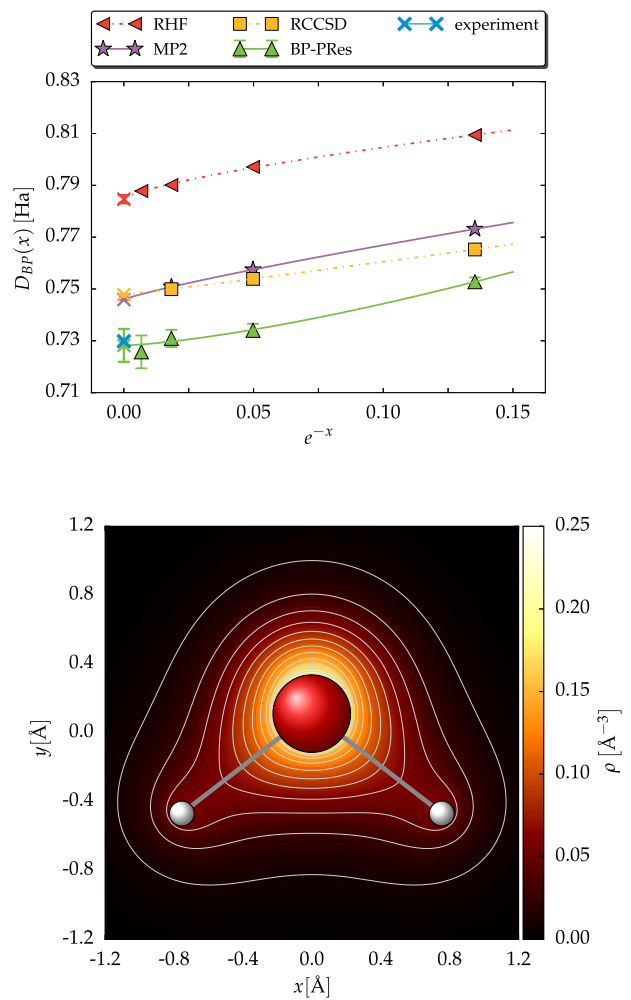}
\caption{
(color online)
Top:    Computed dipole moment of  \chem{H_2O} at the experimental equilibrium geometry, using cc-pV$x$Z bases, with $x=2,3,4,5$, and the extrapolation to the complete basis set limit. 
The corresponding results from RHF, MP2, and RCCSD are also shown.
Bottom: Computed electronic density of \chem{H_2O} by AFQMC along the molecular plane, at 
the cc-pV$5$Z level.
} 
\label{fig:h2o}
\end{figure}

\begin{figure}[t!]
\centering
\includegraphics[width=0.94\columnwidth]{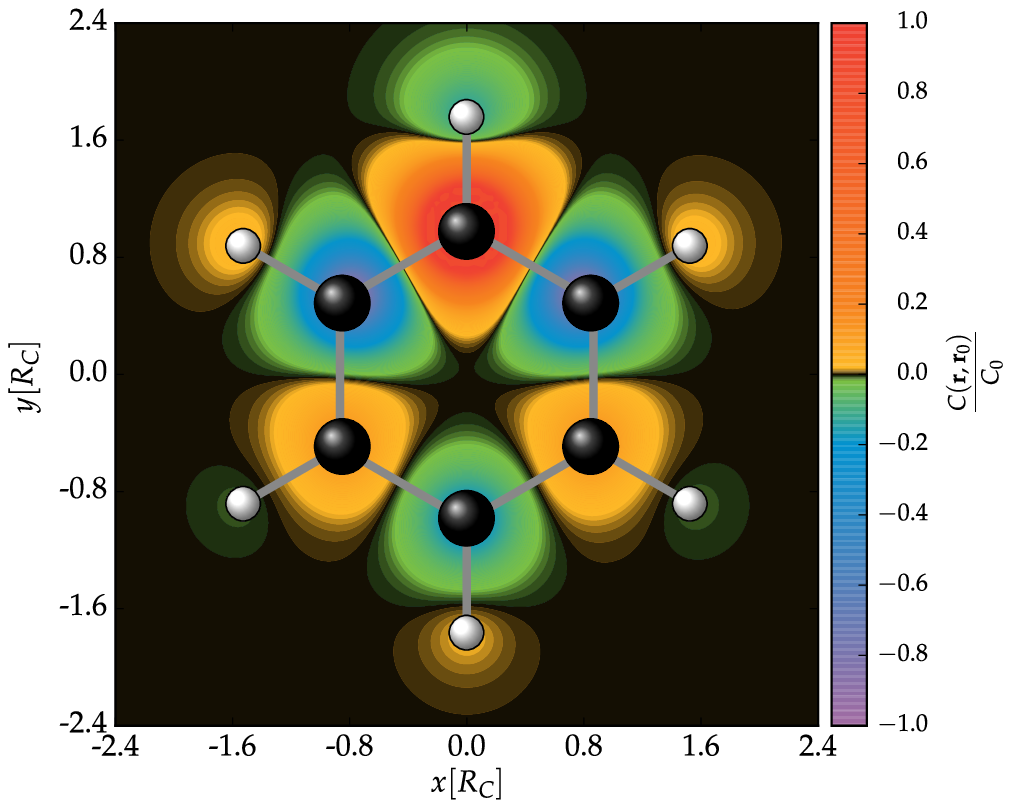}
\caption{
(color online)
Spin-spin correlation function, $C({\bf r},{\bf r_0})$, of benzene at experimental equilibrium geometry,
computed by AFQMC with BP-PRes in the STO-6G basis.
The correlation function is shown for ${\bf r}=(x,y,z_0)$, i.e., in a plane parallel with the molecule 
at a distance $z_0=0.5\,$Bohr above, with the reference spin placed in the same plane directly 
on top of a C atom (the one with the highest $y$ value in the figure). 
$C(x,y)$ is normalized by the magnitude $C_0 = | \min_{x,y} C(x,y) |$ of the most negative value, and 
capped above $C_0$.
} 
\label{fig:benzene}
\end{figure}

We now apply the new BP algorithm, as a first test, to compute ground-state properties of several molecules.
Figure \ref{fig:h2o} shows the computed dipole moment and electronic density of \chem{H_2O} at experimental 
equilibrium geometry \cite{Hoy1979}.
The dipole moment was obtained using cc-pV$x$Z basis sets, with $x=$2,3,4,5. 
Results are extrapolated to the complete basis set limit, $x\to\infty$, using the exponential Ansatz 
$D(x) = \alpha + \beta e^{-\gamma x}$ \cite{Feller1992}.
AFQMC from BP-PRes using a truncated CASSCF(5,7) trial wavefunction yields a dipole moment of 
$\chem{D} = 0.728(3)$ Ha, in agreement with 
the experimental result of  $\chem{D} = 0.7297$ Ha \cite{Lide2003}. The computed result is seen to 
improve appreciably over those from both MP2 and RCCSD.
We also compute the electronic density
\begin{equation}
\rho({\bf{r}}) = \sum_{ij} 2 \, \left(\mathrm{G_{BP}} \right)_{ij} \, \varphi_i({\bf{r}}) \, \varphi_j({\bf{r}}) \,,
\end{equation}
where $\varphi_i({\bf{r}})$ are the molecular orbitals.
The results for water at the cc-pV5Z level are shown in the bottom panel in 
Figure \ref{fig:h2o}.

Back-propagation gives access to all ground-state properties, including the two-body reduced density matrix
and correlation functions of one-body operators. As an example, we compute the spin-spin correlation function
\begin{equation}
\label{eq:benzene}
\begin{split}
 C({\bf{r}},{\bf{r}}_0) &= \sum_{\alpha=1}^3 \frac{ \langle \Psi_0 | \hat{S}_\alpha({\bf{r}}_0) \hat{S}_\alpha({\bf{r}})  | \Psi_0 \rangle } 
              { \langle \Psi_0 | \Psi_0 \rangle } \\
\hat{S}_\alpha({\bf{r}}) &= \sum_{ij,\sigma\tau} (\sigma_\alpha)_{\sigma\tau} \, \varphi_i({\bf{r}}) \, \varphi_j({\bf{r}}) \, \crt{i \sigma} \dst{j\tau}
\end{split}
\end{equation}
where $\sigma_\alpha$ are Pauli matrices and $\varphi_i$, $\varphi_j$ are again molecular orbitals.
In Figure \ref{fig:benzene} we show results for
 \chem{C_6 H_{6}} at experimental equilibrium geometry \cite{Herzberg1966} (specified by the radii $R_C = 1.397$ \AA, $R_{H} = 
2.481$ \AA $\,$ of the \chem{C} and \chem{H} rings, respectively.) 
The figure shows the correlation function in Eq.~ \eqref{eq:benzene} for 
${\bf{r_0}}=(x_0,y_0,z_0)=(0,R_C, 0.5)\,$Bohr, and ${\bf{r}}=(x,y,z_0)$.
This choice corresponds
to placing a reference spin 
above one of the C atoms 
and showing the correlation function in the plane parallel to the molecule. 
Here we used a truncated CASSCF(5,20) state as trial wavefunction.
The alternation between positive and negative values (warm and cold colours) suggests the presence of antiferromagnetic correlations along 
the \chem{C} bonds. Correspondingly a correlation is also present along each \chem{CH} bonds. 

\begin{figure*}[t!]
\centering
\includegraphics[width=0.8\textwidth]{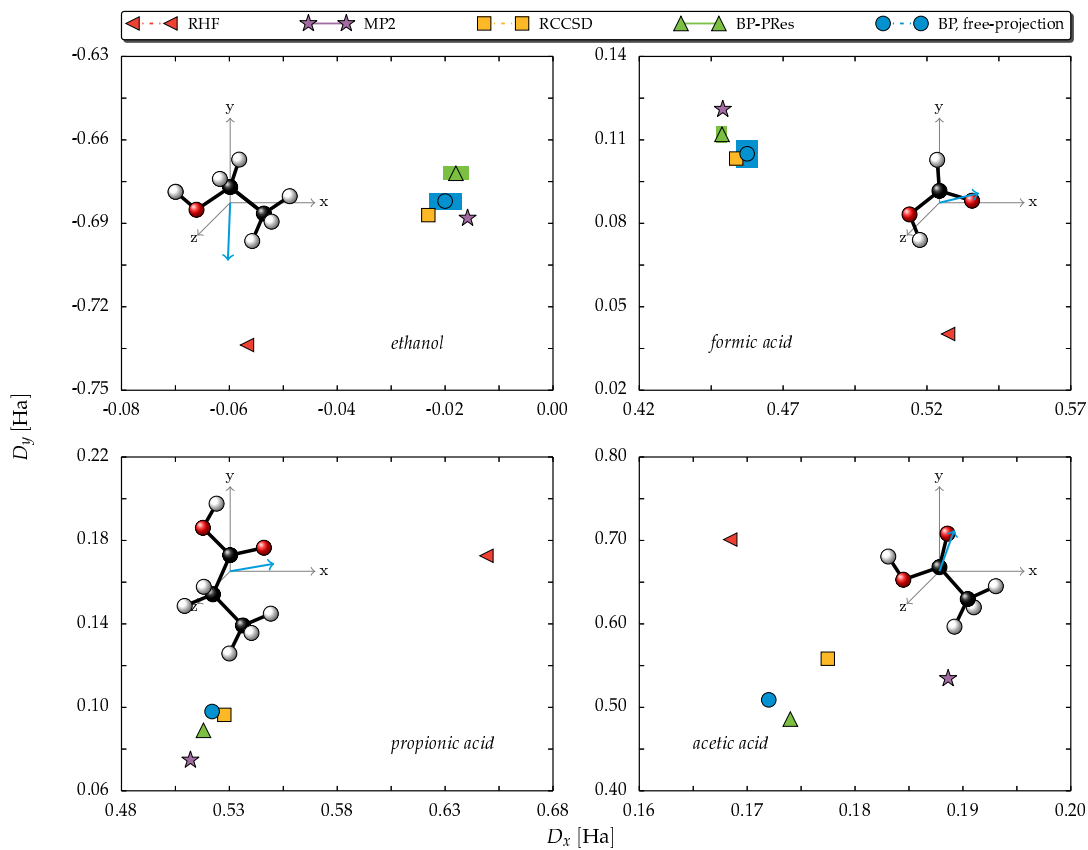}
\caption{
(color online)
Computed dipole moments from AFQMC with BP-PRes for some medium-sized organic molecules
(6-31G*  basis).
Results for
RHF, MP2, and RCCSD are also shown for reference.
Molecular geometries and AFQMC dipoles based on the free-projection scheme are sketched in each panel
(in all cases, in the chosen frame of reference $D_z$ is  equal to or statistically compatible with $0$, 
as expected for these $C_s$ symmetric molecules).
} \label{fig:organic}
\end{figure*}

Finally we compute the dipole moments of a few organic molecules (ethanol \chem{CH_3CH_2 OH}, formic acid \chem{HCOOH}, acetic acid 
\chem{CH_3COOH}, propionic acid \chem{CH_3 CH_2 COOH}) 
in their most stable conformer (trans, trans, syn, $T_t$), at 6-31G$^*$ level. We employ the experimental equilibrium geometries for \chem{CH_3CH_2 OH} 
\cite{Coussan1998} and \chem{HCOOH} \cite{Herzberg1966} and the CCSD, CCSD(T) equilibrium geometries \cite{CCCBDB2016} for \chem{CH_3COOH}
and \chem{CH_3 CH_2 COOH} respectively.
As seen in Figure \ref{fig:organic}, the AFQMC results with BP-PRes  are generally in good agreement with those from RCCSD. 
When a discrepancy is evident, the BP-FP result falls closer to that of BP-PRes. 

\begin{figure}[t!]
\centering
\includegraphics[width=0.94\columnwidth]{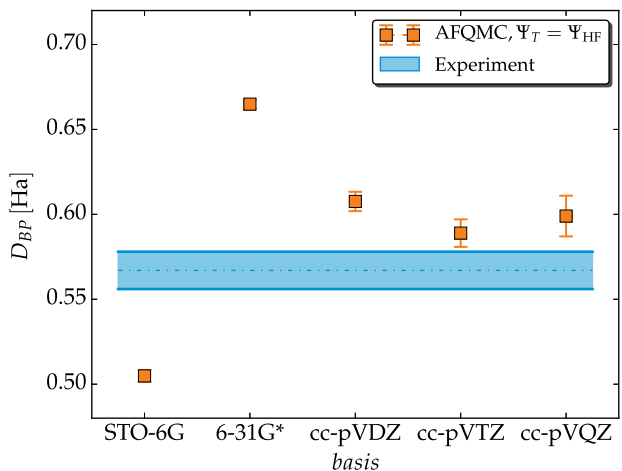}
\caption{
(color online)
Computed dipole moment from AFQMC with BP-PRes, for trans-ethanol at STO-6G to cc-pVQZ level.
Experimental dipole moment is shown for reference.
} \label{fig:ethanol}
\end{figure}

These studies can be carried out beyond the 6-31G$^*$ level, as shown in Figure \ref{fig:ethanol} using \chem{CH_3CH_2 OH} as test case.
AFQMC results are converged to the CBS limit, and lie roughly two joint error bars away from experiment
\cite{CCCBDB2016}.

Given the favorable computational scaling of AFQMC, this is encouraging indication that the method
can potentially provide an accurate description of electronic properties in molecules containing 
hydroxyl, carboxyl and methyl functional groups.

\section{Conclusions}
\label{sec:conclusions}

In  the present work  we have focused on developing the AFQMC method for many-body 
computations in molecules and solids beyond the total energy.  We investigated the use of back-propagation to
compute ground-state observables and correlation functions.
We proposed an algorithm which allows path restoration in the back-propagation (BP-PRes) 
for cases when the phaseless constraint is invoked to control the phase problem.
The algorithm was tested against exact diagonalization and other reference quantum chemistry methods in 
molecules from the first two rows of the periodic table.
We find that significant improvement is achieved with BP-PRes in molecules and solids, where
a phase problem is always present in AFQMC, 
over the simple BP scheme applied in systems with ``only'' a sign problem.

Results are obtained on various quantities including the dipole moment, density matrix, and spin-spin correlation functions. These results 
indicate that AFQMC can become an accurate tool to calculate not only total energy, but all ground-state properties of molecules and real materials. 
We hope that this advance will encourage further development and application of the methodology
to real materials.

\section{Acknowledgments}

\appendix

We thank W. A. Al-Saidi for his contribution at early stages of the project, and 
H. Krakauer and F. Ma for collaborations and discussions. 
M. M. acknowledges Q. Sun and G. K.-L. Chan for help in understanding and using 
PySCF. We acknowledge support by NSF (Grant no. DMR-1409510) and the Simons Foundation. 
Computations were carried out at the Extreme Science and Engineering Discovery 
Environment (XSEDE), which is supported by National Science Foundation grant 
number ACI-1053575, and at the Storm and SciClone Clusters at the College of 
William and Mary.


\end{document}